\documentclass[aps,12pt,eqsecnum,showpacs,nofootinbib]{revtex4}
\usepackage{amsmath}
\usepackage{amsfonts}
\usepackage{amssymb}
\usepackage{color}

\def\bc{\begin{center}}
\def\nno{\nonumber}
\def\ec{\end{center}}
\def\be{\begin{eqnarray}}
\def\ee{\end{eqnarray}}

\newcommand{\omits}[1]{}
\definecolor{dyellow}{rgb}{1.,0.8,.0}
\definecolor{myblue}{rgb}{.1,.1,.7}
\definecolor{dcyan}{rgb}{.0,.6,.6}
\definecolor{dmagenta}{rgb}{0.6,0.0,0.6}
\definecolor{brown}{rgb}{0.6,0.2,0.}
\definecolor{darkblue}{rgb}{.0,.0,0.5}
\definecolor{darkred}{rgb}{0.75,0.0,0.0}
\definecolor{orange}{rgb}{1.,.6,.0}
\definecolor{dorange}{rgb}{0.8,.4,.0}
\definecolor{darkgreen}{rgb}{0.0,0.6,0.0}
\definecolor{purple}{rgb}{.4,.0,.4}

\def\red{\color{red}}

\def\La{\Lambda}
\def\Si{\Sigma}

\def\dl{\delta}
\def\eps{\epsilon}

\def\la{\lambda}

\def\si{\sigma}

\def\d#1#2{\frac{\displaystyle #1}{\displaystyle #2}}

\newcommand\btd{\raise 2pt
\hbox{$\hat\bigtriangledown$}\hskip 1.5pt}
\newcommand\bt{\raise 2pt
\hbox{$\bigtriangledown$}\hskip 1.5pt}

\def\PRD{{\it Phys. Rev.}~{\bf D}}

\begin{document}
\title{On de Sitter Invariant Special Relativity and
Cosmological Constant as Origin of Inertia}

\author{{Han-Ying Guo}$^{1,2,3}$}
\email{hyguo@itp.ac.cn}
\author{{Chao-Guang Huang}$^{2,3}$}
\email{huangcg@mail.ihep.ac.cn}
\author{{Yu Tian}$^{3}$}
\email{ytian@itp.ac.cn}
\author{{Zhan Xu}$^{4}$}
\email{zx-dmp@mail.tsinghua.edu.cn}
\author{{Bin Zhou}$^{2,3}$}
\email{zhoub@itp.ac.cn}

\affiliation{%
${}^1$ China Center of Advanced Science and Technology (World
Lab.), P.O. Box 8730, Beijing
   100080, China,}

\affiliation{%
${}^2$ Institute of High Energy Physics, Chinese Academy of
Sciences, P.O. Box 918-4, Beijing
   100039, China,}

\affiliation{%
${}^3$ Institute of Theoretical Physics,
 Chinese Academy of Sciences,
 P.O.Box 2735, Beijing 100080, China,}

\affiliation{%
${}^4$ Physics Department, Tsinghua University, Beijing
   100084, China.}

\date{June 17, 2004}

\omits{\author{Han-Ying Guo \\
Institute of High Energy Physics, Chinese
Academy of Sciences, P.O. Box 918-4, Beijing 100039, China \\ and
\\ Institute of Theoretical Physics, Chinese Academy of Sciences, P.O.Box 2735, Beijing 100080, China\\ E-mail: \email{hyguo@itp.ac.cn}}

\author{Chao-Guang Huang \\ Institute of High Energy Physics, Chinese
Academy of Sciences, P.O. Box 918-4, Beijing 100039, China \\
E-mail: \email{huangcg@mail.ihep.ac.cn}}

\author{Yu Tian \\ Institute of Theoretical Physics, Chinese Academy of
Sciences, P.O.Box 2735, Beijing 100080, China \\
E-mail: \email{ytian@itp.ac.cn}}

\author{Zhan Xu \\ Physics Department, Tsinghua University, Beijing 100084, China \\
E-mail: \email{zx-dmp@mail.tsinghua.edu.cn}}

\author{Bin Zhou \\ Institute of High Energy Physics, Chinese Academy of
Sciences, P.O. Box 918-4, Beijing 100039, China \\ and \\
Institute of Theoretical Physics, Chinese Academy of Sciences, P.O.Box 2735, Beijing 100080, China \\
E-mail: \email{zhoub@itp.ac.cn}}}

\begin{abstract}
Weakening the Euclidean assumption in special relativity and the
coordinate-independence hypothesis in general relativity for the
de Sitter space, we propose a de Sitter invariant special
relativity with two universal constants of speed $c$ and length
$R$ based on the principle of relativity and the postulate of
universal constants $c$ and $R$ on de Sitter space with Beltrami
metric. We also propose a postulate on the origin of the inertial
motions and inertial systems as a base of the principle of
relativity. We show that the Beltrami-de Sitter space provides
such a model that the origin of inertia should be determined by
the cosmological constant $\Lambda$ if the length $R$ is linked
with $\Lambda$. In addition, via the `gnomonic' projection the
uniform straight-line motion on Beltrami-de Sitter space is linked
with the uniform motion along a great `circle' on de Sitter space
embedded in 5-d Minkowski space.
\end{abstract}



\maketitle

\section{Introduction}

Recent observations show that our universe is accelerated
expanding, asymptotic de Sitter ($dS$) with a positive
cosmological constant $\La$ \cite{90s}, \cite{WMAP}. However,
there are  lots of puzzles related to the $dS$ space. How
to introduce observables? %
What is the statistical origin of the entropy of cosmological
horizon? How to more consistently connect the physical laws in
laboratory scale and in cosmic scale with asymptotic $dS$? There
is no way to get $dS$ space from string theory or M-theory so far,
how to rescue?
And so on so forth.   
These present exciting challenges at both theoretical and
experimental levels and may touch the foundation of physics
\cite{WittenStr}.

In  special relativity (${\cal SR}_c$), one of the breakthroughs
in the last century and of the cornerstones of modern physics, in
addition to the fundamental principles, it is assumed { that the
3-d space (without gravity) is flat as the same as the 3-d space
(without gravity) in Newton mechanics \cite{1905}. Namely, it is
Euclidean. In the large scale of free space, it is less
observation base and should be weakened.  In general relativity
($\cal GR$), another breakthrough and milestone, the 4-d spacetime
is assumed to be not flat in general and the $\La$ term is put in
by hand.  In addition, if one would take the usual understanding
in $\cal GR$ that physics should completely be
coordinate-independent, it might lead to 
the above puzzles and difficulties about $dS$ and asymptotically $dS$ .

On the other hand, as the base of the principle of relativity (PoR) in both Newtonian mechanics and ${\cal SR}_c$ the 
origin of the  inertial motions and inertial coordinate systems,
the origin of inertia 
for short,
together with the origin of inertial mass  are two 
related but different long-standing  problems.
We should also distinguish the origin of inertia from the origin
of the local inertia in $\cal GR$, i.e. the local inertial systems
in the local Minkowski space and the local inertial motions along the
geodesics, respectively. It is an issue of the principle of
equivalence.

As was pointed out \cite{BBR}, in fact, the PoR in the theory of
relativity and the cosmological principle in modern cosmology
  seem to be in conflict. Thus, logically it seems also
hard to explain the origin of inertia
in the theory of relativity, since the all distant stars as a
whole should roughly fit the cosmological principle. The origin of
local inertia seems also the case \cite{weinberg}. Thus, the
origin of inertia seems
still open.

In this paper, we would weaken the Euclidean assumption in ${\cal
SR}_c$ and the coordinate-independence hypothesis in $\cal GR$ for
the $dS$ space  to explore what could happen to it. In addition,
based on recent observations on the dark matter and dark energy or
the cosmological constant as its simplest form, we present a
restatement on the origin of the both inertia and local inertia
and propose a postulate on the origin of inertia for the case
without any matter and dark matter. It is then closely related to
the cosmological principle. It turns out that the ${\cal SR}_c$
can be generalized to a special relativity of $dS$ invariance with
an additional universal constant of length $R$, ${\cal SR}_{c,
R}$. And it provides such a model that the cosmological constant
is just the origin of inertia on $dS$ space via an interesting
relation between the PoR and the cosmological principle on it, if
the length $R$ is linked with $\La$. The analogues approach to the
anti-de Sitter ($AdS$) space can also work.

The ${\cal SR}_{c,R}$ is based on the PoR and the postulate of
invariant universal signal speed $c$ and length $R$ (PoI$_{c,R}$).
If $R$ is taken  as $R^2=3\La^{-1} $, the cosmological constant
appears at principle level and ${\cal SR}_{c,R}$ may also be
denoted as ${\cal SR}_{c,\La}$. Thus, { unlike ${\cal SR}_c$ where
there should be no room for $\La$
and unlike 
$\cal GR$ where there should be no inertial motion with uniform
coordinate velocity}\omits{\footnote{\red Sometimes, an inertial
motion is identified with a geodesic motion in a curved spacetime.
But, a geodesic motion in a generic curved spacetime can only be
regarded as a local inertial motion, at most, because naively, an
inertial motion is a motion of a body which is apart away from
other bodies.}}, there are proper rooms for $\La$ and a kind of
inertial motions with uniform coordinate velocities along straight
lines in the curved spacetimes of constant curvature. We may
define a set of the observable of free particles and generalize
famous Einstein's energy-momentum-mass formula. We may also define
two kinds of simultaneity. The first is directly from PoR and for
the experiments in local laboratories. The second is for the
cosmological observations. It leads to an empty accelerated
expending cosmological model with 3-d space of positive curvature
in the order $\La$. This is an important prediction. Even such a
$dS$ space is empty, but our universe might be so asymptotically.
This is already roughly indicated  by %
$\Omega_T=1.02\pm 0.02$ from WMAP \cite{WMAP}. %

These are in fact  very important properties of $dS$ space
\cite{Lu},\cite{BdS},\cite{Gursey}, which are ignored for long
time in $\cal GR$, might be due to the coordinate-independent
hypothesis approach. Namely, among various metrics of $dS$ spaces,
there is an important one in which $dS$ space is in analog with
Minkowski space. It is the $dS$ space with Beltrami-like metric,
called the Beltrami-de Sitter ($BdS$) space. It is precisely the
Beltrami-like model \cite{beltrami} of a 4-hyperboloid ${\cal
S}_R$ in 5-d Minkowski space, $ BdS \backsimeq {\cal S}_R $. In
$BdS$ space there exist a set of Beltrami coordinate systems
covering $BdS$ patch by patch, and in which  particles and light
signals move along the timelike and null geodesics, respectively,
with {\it constant} coordinate velocities. Therefore, they look
like in free motion in a space without gravity. Thus, the Beltrami
coordinates and observers at these systems should be regarded as
of inertia.

There properties can also be seen as follows. If we start with the
4-d Euclid geometry and weaken the fifth axiom, then there exist
4-d Riemann, Euclid, and Lobachevski
geometries at almost equal footing. For the non-Euclid ones, 
geodesics are in one-to-one correspondence with straight lines in
Beltrami coordinate systems, which are generalizations of the
coordinate systems in Beltrami model of Lobachevski plane
\cite{beltrami}, and under corresponding transformation groups the
systems transform among themselves. Now changing the metric
signature to $-2$, these non-Euclid constant curvature spaces turn
to $dS/AdS$ spaces with the Beltrami metrics, the Euclid one to
Minkowski space $M$, and those straight lines are classified by
timelike, null and spacelike straight world-lines, respectively.
In addition, from projective geometry with an antipodal
identification of, say, $dS$ space $dS/Z_2 \subset RP^4$, it is
also the case except this leading to non-orientability
\cite{Lu},\cite{BdS},\cite{gibbons}. More concretely, one may take
the `gnomonic' projection, which is called sometimes the
`circle-rectilinear' mapping. It maps the great `circle' on
$dS/AdS$ as pseudo-spheres embedded in 5-d Minkowski spaces to the
straight lines on the $dS/AdS$ spaces with the Beltrami metric. Of
course, in order to preserve the orientation, the antipodal
identification should not be taken.

Thus, in analog with ${\cal SR}_{c}$, on $BdS$ space, say, there
should exist such motions and observers of inertia. And PoR should
also be available on it. In addition to the invariant universal
speed, the speed of light $c$, there is another invariant
universal length $R$ as the curvature radii, so the postulate of
invariant of velocity of light should be replaced by the
PoI$_{c,R}$. That is why based on these two principles, the $dS$
invariant special relativity ${\cal SR}_{c,R}$ can be set up.

Furthermore, in view of the ${\cal SR}_{c,R}$, we show that the
$BdS$ space provides such a model, in which the origin of inertia
should be determined by $\La$\omits{\red (with an insight into the
long-standing problem. $\longleftarrow$ Maybe, it is better that
this phrase is not mentioned)}. In addition, via the `gnomonic'
projection
 there is a 
relation between the inertial \omits{uniform straight-line }motion
on $BdS$ space and the uniform motion along a great `circle' on
$dS$ space embedded in a 5-dimensional Minkowski space $M^{1,4}$.

This paper is arranged as follows. In the sections 2-4, we  set up
a framework for ${\cal SR}_{c,R}$ based on the two postulates
\cite{Lu, BdS}. We consider how to introduce a set of observable
for the particles and signals, generalize Einstein's famous
formula and define the intervals, light cone and two kinds of
simultaneity. We show that the 3-cosmic space in the $BdS$ space
is a slightly closed. In the section 5, we present a restatement
on the origin of  both inertia and local inertia based on recent
observations,   propose the postulate on the origin of inertia for
the case without both matter and dark matter and show that the
$BdS$ space is just such a model of $\La$ as the origin of inertia
on it. Finally, we end with a few remarks.

\section{de Sitter invariant special relativity}


We now introduce the two postulates  and set up a framework for
${\cal SR}_{c,R}$ \cite{Lu, BdS}.

 The PoR requires {\it there exist a set of inertial
reference frames, in which the free particles and light signals
move with uniform straight lines, the laws of nature without
gravity are invariant under the transformations among them.} The
\omits{postulate on invariant maximum signal speed and
length}PoI$_{c,R}$ requires {\it there exist two invariant
universal constants --- speed $c$
 and length $R$.}

Owing to Umov, Weyl and Fock \cite{Fock}, it can be proved that
the most general form of the transformations among inertial
coordinate systems $F$ and $F'$%
\be\label{FL}%
{x'}^i=f^i( x^i), \quad x^0=ct, \quad i=0,\cdots, 3,%
\ee
which transform a uniform straight line motion in $F$ \omits{with
eqn.(\ref{uvm}) }to a motion of the same nature in $F'$
\omits{with eqn.(\ref{uvm'}) }are that {\it the four functions
$f^i$ are ratios of linear functions, all with the same
denominators}.

As in ${\cal SR}_c$, \omits{we may naturally require}the PoR
implicates that  the  metric, if it exists, on inertial frame in
spacetime is of signature $\pm 2$ and invariant under a
transformation  group with ten parameters including space-time
`translations' (4), boosts (3) and space rotations (3),
respectively. Thus, due to maximally symmetric space theory
\cite{weinberg},
 the necessary and sufficient condition for 4-d spaces with invariant metric of signature $\pm 2$ under
ten-parameter transformation group is that they are $dS/M/AdS$ of
positive, zero, or negative constant curvature, invariant under
group $SO(1, 4)$, $ISO(1,3)$ or $SO(2, 3)$, respectively. The
PoI$_{c,R}$ requires that there are proper rooms for the constant
$c$ as in  transformations (\ref{FL}) and the invariant  length
$R$, which should be the curvature radii of $dS/AdS$,
respectively.

We now focus on the $dS$ space. It can be regarded as a 4-d
hyperboloid ${\cal S}_R$ embedded in a 5-d Minkowski space with
$\eta_{AB}= {\rm diag}(1, -1, -1, -1, -1)$:
 \be\label{5sphr}%
 {\cal S}_R:  &&\eta^{}_{AB} \xi^A \xi^B= -R^2,%
\\ %
\label{ds2}%
&&ds^2=\eta^{}_{AB} d\xi^A d\xi^B , 
\ee
where 
$A, B=0, \ldots, 4$. Clearly,  Eqs. (\ref{5sphr}) and (\ref{ds2})
are invariant under $dS$ group ${\cal G}_R =SO(1,4)$. Via the
`gnomonic'
projection, ${\cal S}_R$ becomes 
the $BdS$ space, in which there exist Beltrami coordinates
\cite{BdS}\omits{.
 The Beltrami coordinates
are} defined patch by patch. 
For intrinsic geometry of $BdS \simeq{\cal S}_R$ space, there are
at least eight patches $U_{\pm\alpha}:= \{ \xi\in{\cal S}_R :
\xi^\alpha\gtrless 0\}, \alpha=1, \cdots, 4$. In $U_{\pm 4}$, for
instance, the Beltrami coordinates are
\be \label{u4}%
&&x^i|_{U_{\pm 4}} =R {\xi^i}/{\xi^4},\qquad i=0,\cdots, 3;  \\
&&\xi^4|_{U_{\pm 4}}=({\xi ^0}^2-\sum _{a=1}^{3}{\xi ^a}^2+ R^2
)^{1/2} \gtrless 0,
\ee%
 which are like locally the inhomogeneous coordinates in
projective geometry but {\it without the antipodal
identification}. In the patches $U_{\pm a}$, $a=1,2,3$,
\begin{equation}  %
y^{j'}|_{U_{\pm a}}=R{\xi^{j'}} /{\xi^a},\quad
j'=0,\cdots,\hat{a}\cdots,4; \quad \xi^{a}\neq 0,
\end{equation}
where $\hat{a}$ means omission of $a$. It is important that all
transition functions in  intersections are of ${\cal G}_R$ in the
type (\ref{FL}). For example, in $U_4\bigcap U_3$, the transition
function $T_{4,3} =\xi^3/\xi^4=x^3/R=R/y^4 \in {\cal G}_R$ so that
$x^i=T_{4,3}y^{i'} $.

In each patch, there are condition and Beltrami metric
\begin{eqnarray}\label{domain}
\sigma(x)&=&\sigma(x,x):=1-R^{-2}
\eta_{ij}x^i x^j>0,\\
\label{bhl} ds^2&=&[\eta_{ij}\sigma(x)^{-1}+ R^{-2}
\eta_{ik}\eta_{jl}x^k x^l \sigma(x)^{-2}]dx^i dx^j.
\end{eqnarray}
 Under fractional linear
transformations of type (\ref{FL}), which are transitive in ${\cal G}_R$ sending $A(a^i)$ to the origin,%
\begin{equation}\label{G}
\begin{array}{l}
x^i\rightarrow \tilde{x}^i=\pm\sigma(a)^{1/2}\sigma(a,x)^{-1}(x^j-a^j)D_j^i,\\
D_j^i=L_j^i+{ R^{-2}}%
\eta_{jk}a^k a^l (\sigma(a)+\sigma(a)^{1/2})^{-1}L_l^i,\\
L:=(L_j^i)_{i,j=0,\cdots,3}\in SO(1,3),
\end{array}\end{equation}
where $\eta_{ij}%
={\rm diag} (1, -1,-1,-1)$ in $U_{\pm\alpha}$, {\it  Eqs.
(\ref{domain}) and (\ref{bhl}) are invariant. The inertial frames
and inertial
motions transform among themselves, respectively.} Note that Eqs. (\ref{domain})-(\ref{G}) are defined on 
 patch by patch. \omits{ This is, in fact, a cornerstone for
the SR-type principle. }$\sigma(x)=0$ is the projective boundary
of $BdS$, denoted by $\partial(BdS)$.

For two separate events $A(a^i)$ and $X(x^i)$ in $BdS$,
\be\label{lcone0} %
{\Delta}_R^2(a, x) = R^2
[\sigma^{-1}(a)\sigma^{-1}(x)\sigma^2(a,x)-1]
\ee %
is invariant under ${\cal G}_R$. Thus, the interval between $A$
and $B$ is timelike, null, or spacelike,
respectively, according to%
\begin{equation}\label{lcone}%
\Delta_R^2(a, b)\gtreqqless 0.%
\end{equation}

The proper length of timelike or spacelike between $A$ and $B$ are
integral of ${\cal I} ds$ over the geodesic segment
$\overline{AB}$:
\be \label{AB1}%
S_{t-like}(a, b)&=&R \sinh^{-1} (|\Delta(a,b)|/R), \\
\label{AB1sl} S_{s-like}(a,b)&=& R \arcsin (|\Delta(a,b)|/R),
\ee
where ${\cal I}=1, -i$ for timelike or spacelike, respectively.
Note that there exist such kind of  pairs ($A$, $B$) that there is
no geodesic segment $\overline{AB}$ to connect them. We will
explain this issue elsewhere.

The light-cone at $A$ with running points $X$ is
\be \label{nullcone} %
{\cal F}_{R}:= R
\{\sigma(a,x) \mp [\sigma(a)\sigma(x)]^{1/2}\}=0.%
 \ee%
It satisfies the null-hypersurface condition.\omits{
 \be\label{Heqn}%
\left . g^{ij}\frac{\partial f}{\partial x^i}\frac{\partial
f}{\partial x^j}\right |_{f=0}=0, %
\ee
where $g^{ij}=\sigma(x)(\eta^{ij}-R^{-2} x^i x^j)$ inverse of
(\ref{bhl}).} %
At the origin %
$a^i=0$, the light cone becomes a Minkowski one and 
$c$ is numerically the speed of light in the vacuum.
\section{Inertial motion, observables and Beltrami simultaneity}


It is easy to check that on $BdS$ space the geodesics are
Lobachevski-like straight lines. In fact, the `gnomonic'
projection provides such an intuitive\omits{onistic} picture.

\omits{According to the PoR, }A particle with mass $m_{\Lambda}$
moves along a timelike geodesic, \omits{}%
which
has the first integration
\begin{equation}\label{pi}
\frac{dp^i}{ds}=0, \quad p^i:=m_{\Lambda
}\sigma(x)^{-1}\frac{dx^i}{ds}=C^i={\rm const.}
\end{equation}
This implies \omits{its coordinate velocity components are
constants and }under the initial condition
\[
x^i(s=0)=b^i, \qquad \frac{dx^i}{ds}(s=0)=c^i
\]
with the constraint $ g_{ij}(b)c^i c^j=1, $ \omits{a new parameter
$w = w(s)$ can be chosen such that }the geodesic is just a
straight world-line
\begin{equation}
x^i(w)=c^iw+b^i,
\end{equation}
where $w = w(s)$ is given by \omits{This property is in analog
with the straight line in the Beltrami model of Lobachevski plane.
The parameter $w$ can be integrated out,}
\be \label{w1}
w(s) = \begin{cases}
R e^{\mp s/R}\sinh \d s R,   &    \eta_{ij}\,c^i c^j = 0, \medskip \cr
\d {R\sinh \frac s R} {\frac {\eta_{ij}\,c^i b^j}{R\sigma(b)}\sinh \frac s R
    + \cosh \frac s R},  & \eta_{ij}\,c^i c^j \neq  0.
\end{cases}
\ee

Similarly, a light signal moves inertially
along a null geodesic, which 
still has the first integration
\begin{equation}
\sigma ^{-1}(x)\frac{dx^i}{d\la }={\rm constant},
\end{equation}
where $\la $ is an affine parameter. Under the constraint $
g_{ij}(b)\,c^ic^j=0$ and  initial condition
\be %
x^i(\la =0)=b^i,  \qquad \d {dx^i}{d\la }(\la =0)=c^i,%
\ee
the null geodesic can be expressed as a straight line
\be x^i = c^i w(\la) +b^i, \ee
where
\be \label{w2}%
w(\tau) = \begin{cases} {~ \la}, & \eta _{ij}\,c^ic^j =0, \cr - \d
{ R^2\sigma (b)}{|\eta _{ij}c^ic^j|}\left( \d 1{\la +\la _0}-\d
1{\la _0}\right), & \eta _{ij}\,c^ic^j \neq 0,
\end{cases}
\ee
with
\begin{equation}\nonumber
\la _0=\sqrt{\frac{ R^2 \sigma (b)}{|\eta _{ij}c^ic^j|}}.
\end{equation}

Thus, the  motions of both free particles and light signals are
indeed inertia, i.e. 
 the coordinate velocity
components  are constants, respectively:
\begin{equation}\label{vi}
\frac{dx^a}{dt}=v^a;\quad \frac{d^2x^a}{dt^2}=0;\quad a=1,2,3.
\end{equation}
 \omits{Of course, this makes sense only if the Beltrami
coordinate system is of physical meaning as inertial-type.} Note
that these properties are well defined patch by patch. Conversely, 
all  straight lines are geodesics.

In ${\cal GR}$, however, these properties are ignored might be due
to the coordinate-independence hypothesis.

Now we define  sets of the observable for free particles and
signals. From (\ref{pi}), it is natural to define the conservative
quantities $p^i$ as the 4-momentum of a free particle with mass
$m_{\Lambda}$ and its zeroth component as the energy. Note that it
is no longer a 4-vector rather a pseudo 4-vector. Furthermore, for
the particle a set of conserved quantities $L^{ij}$ may also be
defined
\begin{equation}\label{angular4}
L^{ij}=x^ip^j-x^jp^i;\quad \frac{dL^{ij}}{ds}=0.
\end{equation}
These may be called the 4-angular-momentum, which is a pseudo
anti-symmetric tensor. Further, $p^i$ and $L^{ij}$ constitute a
conserved 5-d angular momentum in ${\cal S}_R$
\begin{equation}\label{angular5}
{\cal L}^{AB}:=m_{\Lambda
}(\xi^A\frac{d\xi^B}{ds}-\xi^B\frac{d\xi^A}{ds}); \quad
\frac{d{\cal L}^{AB}}{ds}=0.
\end{equation}
Thus, 
(\ref{angular5}) defines {\it a kind of  the uniform motions along
 great `circles' hidden on ${\cal S}_\La $ that related to the
inertial motions on $BdS$ space via the `gnomonic' projection
without antipodal identifications and vice versa.}

 For such a kind of free particles, there is a {\it generalized
Einstein's formula including 4-angular momentum}:%
\begin{equation}\label{eml}
-\frac{1}{2R^2}{\cal L}^{AB}{\cal L}_{AB}=E^2-{p_bp^b}-
\frac{1}{2R^2} L^{ij}L_{ij}=m_{\Lambda}^2,
\end{equation}
where ${\cal L}_{AB}=\eta_{AC}\eta_{BD}{\cal L}^{CD}$,
$p_a=\delta_{ab}p^b$,
$L_{ij}=\eta_{ik}\eta_{jl}L^{kl}$. Here  $m_{\Lambda }$ %
the {\it $dS$ invariant inertial mass for a free particle}
well-defined along with the energy, momentum, boost and angular
momentum at classical level. Further, $m^2_{\Lambda}$ is the
eigenvalue of the first Casimir operator of the $dS$ algebra.
Thus, ${\cal SR}_{c,R}$ offers {\it a $dS$ invariant definition
and classification of the inertial mass}.
In addition, spin 
can also be well defined and related to the eigenvalue of the
second
Casimir operator. 
\omits{}In fact, the generators forming an $so(1,4)$
algebra of (\ref{G}) 
 read in $BdS$ space  %
\begin{equation}\label{generator}
\begin{array}{l}
  \mathbf{P}_i =(\delta_i^j-R^{-2}x_i x^j) \partial_j, \quad  x_i:=\eta_{ij}x^j,\\
  \mathbf{L}_{ij} = x_i \mathbf{P}_j - x_j \mathbf{P}_i
  = x_i \partial_j - x_j \partial_i \in so(1,3),
\end{array}
\end{equation}\omits{
They form an $so(1,4)$ algebra
\begin{eqnarray}
  [ \mathbf{P}_i, \mathbf{P}_j ] &=& R^{-2} \mathbf{L}_{ij} \nno\\
  {[} \mathbf{L}_{ij},\mathbf{P}_k {]} &=&
    \eta_{jk} \mathbf{P}_i - \eta_{ik} \mathbf{P}_j
\label{so(1,4)}\\
  {[} \mathbf{L}_{ij},\mathbf{L}_{kl} {]} &=&
    \eta_{jk} \mathbf{L}_{il}
  - \eta_{jl} \mathbf{L}_{ik}
  + \eta_{il} \mathbf{L}_{jk}
  - \eta_{ik} \mathbf{L}_{jl}. \nno
\end{eqnarray} }\omits{

Thus, ${\cal SR}_{c, R}$ offers a consistent way to define the
observables for free particles. These issues
significantly confirm  that the motion of a free particle 
is  inertial, the Beltrami coordinate systems and corresponding
observer  of the system are all inertial.}with two Casimir
operators 
\be \label{C1} \mathbf{C}_1 &=& \mathbf{P}_i \mathbf{P}^i -\frac 1
2 R^{-2} \mathbf{L}_{ij} \mathbf{L}^{ij},\\
\mathbf{C}_2 &=& \mathbf{S}_i \mathbf{S}^i -R^{-2} \mathbf{W}^2,
\ee
where $\mathbf{P}^i = \eta^{ij} \mathbf{P}_j,
\mathbf{L}^{ij}=\eta^{ik} \eta^{jl} \mathbf{L}_{kl}$,
$\mathbf{S}_i = \frac 1 2 \eps_{ijkl} \mathbf{P}^j
\mathbf{L}^{kl}$, $\mathbf{S}^i =\eta^{ij} \mathbf{S}_j$,
$\mathbf{W} =\frac 1 8 \eps_{ijkl} \mathbf{L}^{ij}
\mathbf{L}^{kl}$. \omits{ In addition, spin can also be defined
through the eigenvalue of the second Casimir operator \be
\mathbf{C}_2 = \mathbf{S}_i \mathbf{S}^i -R^{-2} \mathbf{W}^2 \ee}
\omits{For such a kind of free particles, there is a {\it
generalized
Einstein's famous formula}:%
\begin{equation}\label{eml}
-\frac{1}{2R^2}{\cal L}^{AB}{\cal L}_{AB}=E^2-{\bf P}\,^2-
\frac{1}{2R^2}{\bf L}^2=m_{\Lambda}^2,
\end{equation}
where ${\cal L}_{AB}=\eta_{AC}\eta_{BD}{\cal L}^{CD}$, $m_\Lambda
\geq 0$ introduced above should be the {\it $dS$ invariant
inertial mass} for a free
 particle. It is well defined together with the energy, momentum and angular momentum at classical level.

It can further be shown that $m_{\Lambda}$ is the eigenvalue of
the first Casimir operator of the $dS$ group ${\cal G}_R$
\cite{Gursey}.
In addition, spin 
can also be well defined as it was done in the relativistic
quantum mechanics in Minkowski space. \omits{}}

Thus, ${\cal SR}_{c, R}$ offers a consistent way to define a set
of the observable for free particles. These issues
significantly confirm  that the motion of a free particle 
is  inertial, the Beltrami coordinate systems and corresponding
observer  of the system are all inertial as well.

As in ${\cal SR}_{c}$, in order to make sense  of inertial motions
and observables practically, one should define simultaneity and
take space-time measurements. In ${\cal SR}_{c}$, due to the PoR,
Minkowski coordinates have measurement significance. Namely, the
difference in time coordinate stands for the time interval, and
the difference in spatial coordinate stands for the spatial
distance. Similar to ${\cal SR}_{c}$, one can define that two
events $A$ and $B$ are simultaneous if and only if the Beltrami
time coordinate $x^0$ for the two events
are same, 
\be %
a^0:=x^0(A) =x^0(B)=:b^0. %
\ee
It is called the {\it Beltrami simultaneity} and with respect to
it that free particles move inertially. The Beltrami simultaneity
defines a 3+1 decomposition of spacetime
\be ds^2 =  N^2 (dx^0)^2 - h_{ab} \left (dx^a+N^a dx^0 \right )
\left (dx^b+N^b dx^0 \right ) %
\ee
with the lapse function, shift vector, and induced 3-geometry on
3-hypersurface $\Si_c$ in one coordinate patch.
\begin{eqnarray}
& & N=\{\si_{\Si_c}(x)[1-(x^0 /R)^2]\}^{-1/2}, \nonumber \\%
& & N^a=x^0 x^a[ R^2-(x^0)^2]^{-1},
 \\
& & h_{ab}=\dl_{ab} \si_{\Si_c}^{-1}(x)-{ [R\si_{\Si_c}(x)]^{-2}
\dl_{ac} \dl_{bd}}x^c x^d ,\nonumber
\end{eqnarray}
respectively, where $\si_{\Si_c}(x)=1-(x^0{/R})^2 + {\dl_{ab}x^a
x^b /R^2}$,  $\dl_{ab}$ is the Kronecker $\dl$-symbol,
$a,b=1,2,3$.  In particular, at $x^0=0$, $\si_{\Si_c}(x)=1+{
\dl_{ab} x^a x^b/R^2}$. In a vicinity of the origin of Beltrami
coordinate system in one patch, 3-hypersurface $\Si_c$ is
isomorphic to a 3-sphere. \omits{acts as a
Cauchy surface.}For the $x^0\neq 0$, it is also the case.

The Beltrami simultaneity defines the laboratory time in one
patch. In the spirit of ${\cal SR}_{c}$ and due to the PoR in
${\cal SR}_{c,R}$, there are definite relations between the
Beltrami coordinates and the standard clocks and  rulers in
laboratory in such a manner that measure the time of a process or
the size of an object, one may just need to compare with Beltrami
coordinates together with their relations with the standard clocks
and  rulers.

\section{Proper-time simultaneity %
and slightly closed cosmic space}

\omits{\it Proper-time simultaneity %
and closed cosmic space.}

It should be emphasized that there is another simultaneity in
${\cal SR}_{c, R}$ and it is directly related to the cosmological
principle. It
is {\it proper-time simultaneity} 
with
respect to 
a clock rest at spatial origin of the Beltrami coordinate system.

The proper time $\tau=\tau_{\Lambda}$ of a rest clock on the time
axis of Beltrami coordinate system, $\{x^a=0\}$, reads
\begin{eqnarray}\label{ptime}
\tau=\tau_{\Lambda}=R\sinh^{-1}(R^{-1}\sigma^{-1/2}(x) x^0).
\end{eqnarray}
Therefore, we can define that the events are simultaneous with
respect to it, then these events are co-moving with the clock, if
and only if
\begin{equation}\label{smlt}
x^0\sigma^{-1/2}(x,x)=\xi^0:=R \sinh(R^{-1}\tau)={\rm constant}.
\end{equation}
The line-element %
on the simultaneous 3-d
hypersurface, denoted by ${\Sigma_\tau}$, can be defined as
\begin{equation}\label{dl}
dl^2=-ds^2_{\Sigma_\tau},%
\end{equation}
where
\be \begin{array}{l}\label{spacelike}
ds^2_{\Sigma_\tau} = R_{\Si_\tau}^2%
dl_{{\Sigma_\tau} 0}^2, \\
R_{\Sigma_\tau}^2%
:=\sigma^{-1}(x,x)\sigma_{\Sigma_\tau}(x,x)%
=1+ (\xi^0/R)^2,\\
\sigma_{\Sigma_\tau}(x,x):=1+R^{-2}\delta_{ab}x^a x^b
>0, \\
dl_{{\Sigma_\tau} 0}^2:={
\{\delta_{ab}\sigma_{\Sigma_\tau}^{-1}(x)
-[R\sigma_{\Sigma_\tau}(x)]^{-2}\delta_{ac}\delta_{bd}x^c x^d\}}
 dx^a dx^b.
\end{array}\ee

The relation of this simultaneity with the cosmological principle
can be seen as follows. In fact,
 it is significant that if $\tau_{\Lambda}$ is taken as
a ``cosmic time", the Beltrami metric (\ref{bhl}) becomes a
Robertson-Walker-like metric with a positive spatial curvature and
the simultaneity is globally defined in whole $BdS$ space
\begin{equation}\label{dsRW}
ds^2=d\tau^2-dl^2=d\tau^2-\cosh^2( R^{-1}\tau) dl_{{\Sigma_\tau} 0}^2.%
\end{equation}
This  shows that {\it the 3-d cosmic space is $S^3$ rather than
flat. The deviation from the flatness is of order $\Lambda$.} Our
universe should be asymptotically so.

The two definitions of simultaneity do make sense in different
kinds of measurements. The first concerns the measurements in a
laboratory and is related to the PoR and PoI$_{c,R}$ of ${\cal
SR}_{c, R}$. The second concerns with cosmological effects.
Furthermore, the relation between the Beltrami metric with
coordinate time $x^0$ and its RW-like counterpart with cosmic time
$\tau$ links the PoR and cosmological principle. It is very
meaningful. The prediction that the spatial closeness of the
universe asymptotically in the order of $\La$ is  different from
the standard cosmological model with flatness or with a free
parameter $k$.

\section{The postulate on the origin of inertia and $\La$ as the origin of inertia in ${\cal
SR}_{c,
R}$}
\omits{\it 2. The postulate on origin}

\omits{Recent observations in precision cosmology show that our
universe is accelerated
expanding, asymptotic $dS$ dominated by the dark sector 
\cite{90s, WMAP} and may prefer  the simplest form of the dark
energy, a positive cosmological constant $\La$ \cite{Wang04}.
However, there are lots of puzzles related to the $dS$ and
asymptotic $dS$ spaces that present exciting challenges at both
theoretical and experimental levels and may touch the foundation
of physics \cite{Peebles, Witten, Str}. \omits{Keeping these
puzzles in mind, it might be better to start with a re-exam into
the foundation of physics.}}

Based upon recent observations, it is clear that in such a
universe the problem of the origin of inertia and its local
version should be restated in the
sprit of Riemann, 
Mach 
and Einstein: {\it The origin of inertia and the origin of local
inertia should be\omits{Thus, we propose a postulate
on the origin of inertia and its local version: 
or the new version of Mach's principle (NMP): {\it Both the origin
of inertia and local inertia should be}} dominatingly determined
by  
the dark matter and dark energy or the
cosmological constant 
along with a little contribution from the distant stars and other
regular matter and radiation.} Since all of them as a whole
support the cosmological principle, if the statement might be true, it should be 
closely related to the cosmological principle except the cases
that some individual star/matter is as a local gravitational
source and so on.
On the other hand, if this statement could make sense, 
for the empty $dS$ space without either matter or dark matter, the
$\La$ might contribute to the origin of inertia rather than its
local version. Thus, we should also have such a postulate on
inertia: {\it If in $dS$ space there could exist the inertial
motions and inertial systems, they should be determined
by the 
cosmological constant.}   
\omits{In fact, if there postulate on the origin of inertia could
make sense, there
should exist inertial motions and systems in some $dS$ space. without 
the dark matter and distant stars.}


\omits{Recently, \omits{In view of the $dS$ invariant ${\cal
SR}_{c,
R}$}it is shown \cite{BdS, Lu} that }As was mentioned in the last sections, on the 
$BdS$ space
there is such a relation between the PoR and 
the cosmological principle. Thus, it indicates that $BdS$ space
may provide such a model on $\La$ as the origin of inertia. In fact, 
on $BdS$ space two kinds of simultaneity just link the two
principles. Namely, as was shown above, the Beltrami simultaneity
is directly from the PoR for the experiments in
the local laboratories. 
The proper time simultaneity is for the cosmological observations.
If the proper time is taken as a cosmic time, it
leads $BdS$ space as an `empty' accelerated expending cosmological model 
and 3-d space of positive curvature
in the order $\La$. 

Importantly, it should be pointed out that the inertial motions
and the inertial coordinate systems of the PoR on $BdS$ space
should be determined by $\La$ via the cosmological principle. What
is needed to do is changing the timing from the cosmic time $\tau$
to the Beltrami time $x^0=ct$ and vice versa. Namely, if the
comoving observers on (\ref{dsRW}) would change the time
measurement from the cosmic time $\tau$ to the Beltrami time $x^0$
according to the relation (\ref{ptime}), they should become a kind
of inertial
observers 
and vice versa.

Thus,  the ${\cal SR}_{c, R}$ on $BdS$ space provides such a model
that the origin of inertia is just $\La$ and it seems to be
 more complete and consistent in logic within the PoR, postulate on inertia\omits{restricted
in  $\La$} and cosmological principle. 
\omits{\it the $BdS$ space 
is such a model that the postulate on inertia 
directly leads to the inertial motions and inertial systems for
the principle of relativity.}


\section{Conclusions}
\omits{\it Conclusions.}

Weakening the Euclidean assumption in ${\cal SR}_c$ and the
coordinate-independence hypothesis in $\cal GR$ for the $dS$
space, we have set up the $dS$ invariant special relativity with
an invariant length $R$ in addition to $c$, ${\cal SR}_{c,
\La>0}$. It is based on the PoR and PoI$_{c,R}$. Similarly, ${\cal
SR}_{c, \La <0}$ with $AdS$ invariance can also be set up. The
Beltrami coordinates, which are like locally inhomogeneous
projective coordinates but without the antipodal identification,
are inertial, the test particles and signals move inertially along
the timelike, null straight world lines, respectively.  There are
also $dS$ invariant definitions and classifications of the
inertial mass and spin of the free particles and fields.
Einstein's energy-momentum-mass formula is generalized. Intervals,
light cone and two kinds of simultaneity are defined. There is
also an interesting relation between the uniform motion along a
great `circle' hidden on ${\cal S}_\La$ and the inertial motion on
$BdS$ space. \omits{A set of classical observable for them can be
well defined. }

It is important the relation between the Beltrami metric and the
RW-like metric. It links the PoR and cosmological principle,
relates the coordinate time $x^0$ in the laboratory and the cosmic
time $\tau$ in the cosmic scale. This also predicts that
asymptotically the 3-d cosmic space is  slightly closed in order
of $\La$. Hopefully, this prediction should be confirmed by the
further data from WMAP in large scale.

We have also proposed a statement on the origin of all kinds of
inertial motions and inertial systems based upon the recent
observations on the dark sector in the cosmic scale and a
postulate on the origin of inertia without any matter and dark
matter for $dS$ space. We show that it does make sense that the
$\La$ provides the origin of inertia in the $BdS$ space. In
addition, it can also be shown that the Newton-Hooke space-times
${\cal NH}_\pm$ contracted from Beltrami-$dS/AdS$ are also such
kind of models with the origin of inertia.

Most properties here are in analog with ${\cal SR}_{c}$ except
there is a proper room of $R$ or $\La$, which leads to those
amazing aspects, and coincide with ${\cal SR}_{c}$ if $\La\to 0$,
i.e. the ${\cal SR}_{c}$ is ${\cal SR}_{c, \La\to 0}$. However,
all amazing aspects of the $dS$ invariant ${\cal SR}_{c, R}$
disappear under such a limitation.

All local experiments would allow there might exist three theories
of special relativity at almost equal footing, i.e. ${\cal SR}_{c,
\Lambda>0}$, ${\cal SR}_{c, \Lambda=0}$, and ${\cal SR}_{c,
\Lambda <0}$ with $dS$, Poincar\'e, and $AdS$ invariance,
respectively. Recent observations in cosmic scale show that there
should be a positive cosmological constant. Therefore, the $dS$
invariant special relativity, ${\cal SR}_{c, \Lambda
>0}$, is more reasonable candidate for a starting point if one
would modify and apply  the relativistical physics to our
expanding asymptotic $dS$ universe as a whole in a more consistent
manner.

In summery, if Galileo Galilei could arrange a large spaceship
voyage, his friends might not need to shut themselves up `in the
main cabin below decks'\cite{GG}.  During the voyage they could
even carry on, with a sensitive microwave telescope and other
instruments, the measurement of the spaceship drift with respect
to the CMB and to explore how much of the $\La$ should contribute
to the origin of inertial motion as well as to the origin of the
hidden uniform motion along a great `circle' in an asymptotic
$dS$-cosmos.

\begin{acknowledgments}
We would like to thank Professors Q. K. Lu, J. Z. Pan, Y.S. Wu and
C.J. Zhu for valuable discussions and comments. We are also
grateful to Prof. G.W. Gibbons for his comments on related works.
This work is partly supported by NSFC under Grants Nos. 90103004,
10175070, 10375087, 10347148.
\end{acknowledgments}

\end{document}